\documentclass[prl,twocolumn]{revtex4}

\usepackage{graphicx}
\usepackage{dcolumn}
\usepackage{bm}

\begin{document}

\title{On the nature of the 5f states in $\delta$-plutonium}

\author{J.M. Wills}
\affiliation{Theoretical Division, Los Alamos National Laboratory}
\author{O. Eriksson}
\affiliation{ Department of Physics, Uppsala University, Box 530, Sweden }
\author{A. Delin}
\affiliation{ ICTP, Strada Costiera 11, 34100 Trieste, Italy}
\author{P.H.Andersson}
\affiliation{ Swedish Defense Research Agency (FOI), 17290 Stockholm, Sweden}
\author{J.J. Joyce, T. Durakiewicz, M.T. Butterfield and A.J. Arko }
\affiliation{ Condensed Matter and Thermal Physics Group, Los Alamos National Laboratory}
\author{D.P. Moore and L.A. Morales}
\affiliation{ Nuclear Materials Technology Division, Los Alamos National Laboratory}

\date{March 26, 2003}

\begin{abstract}
We present a theoretical model of the electronic structure of
$\delta$-Pu that is consistent with many of the electronic
structure related properties of this complex metal. In particular
we show that the theory is capable of reproducing the valence band
photoelectron spectrum of $\delta$-Pu. We report new experimental
photoelectron spectra at several photon energies and present
evidence that the electronic structure of $\delta$-Pu is unique
among the elements, involving a 5f shell with four 5f electrons in
a localized multiplet, hybridizing with valence states, and
approximately one 5f electron forming a completely delocalized
band state.

\end{abstract}

\maketitle

The phase diagram of plutonium (Pu) is extremely
complex\cite{young}. In fact, it has the most complex phase
diagram of all metals, both with regard to the crystal structures
and the number of different phases. Of the different allotropes of
Pu, it is evident that the $\delta$-phase is the most
controversial and anomalous one, and consequently most of the
recent experimental and theoretical work has focussed on this
allotrope.
From an experimental point of view, the reactivity and
radioactivity of Pu, and the complexity of the phase diagram, make
it exceedingly complicated to collect high-quality data. In this
report, we present new and conclusive experimental valence
photoemission spectra taken under highly controlled conditions for
$\delta$-Pu.
The difficulty from a theoretical point of view is to describe, within a unified model,
the anomalously large volume of  $\delta$-Pu,
its energetic proximity to the $\alpha$-phase,
the absence of magnetic order, and the structure of the photoemission spectrum.
So far, none of the proposed theoretical models have been able to do this.

We present in this paper new, high-quality experimental photoemission data for this material
and describe a theoretical model that is consistent with this data
as well as the experimentally observed volume, energy, and lack of magnetic order
in $\delta$-Pu.
Our theory is based on
a partitioning of the electron density into localized and delocalized parts,
minimizing the total energy
(including a correlation energy associated with localization)
with respect to the partitioning.
The minimum energy of $\delta$-Pu is found at a lattice constant close to the experimental volume, with
four 5f electrons localized in an atomic singlet, leaving one itinerant 5f electron
contributing to the chemical bond.

The structural properties of the low-temperature $\alpha$-phase,
which crystallizes in a monoclinic structure, are in stark
contrast to that of the high-temperature, cubic (fcc)
$\delta$-phase, stable between $319^\circ$ and $451^\circ$C. The
volume of this phase is $\sim$25\% greater than the
$\alpha$-volume, and the $\alpha \to \delta$ phase transition is
hence accompanied by an unusually large volume expansion. However,
the addition of only a few percent of, say, gallium (or several
other trivalent metals) stabilizes the $\delta$-phase at low
temperatures, which implies that the total energies of the
$\delta$- and $\alpha$-phases must be nearly equal. In addition to
the very different densities and structural properties, the
$\alpha$- and $\delta$-phases have very different thermal
expansion coefficients \cite{perterm} and mechanical
properties\cite{pudata}.

The vastly different structural and mechanical behaviour of the $\delta$- and $\alpha$-phases
suggest that their
electronic structures must be qualitatively different, and there have been several theoretical
efforts trying to explain these differences.
Several theoretical papers
report on first principles calculations using the local density
approximation (LDA)\cite{brooks,solovyev,soderlind}, LDA
with orbital polarization\cite{soderlind-2}, the
LDA+$U$ approach\cite{bouchet,price,savrasov-LDAU}, LDA+$U$ combined with dynamical mean
field theory (DMFT)\cite{savrasov-DMFT}, self interaction
correction\cite{petit}, the spin generalized gradient approximation (GGA)\cite{soderlind-4}
and the mixed level model (MLM)\cite{eriksson}.
Many of the theoretical models are in disagreement with each other, in terms of how one
should understand the electronic structure of the different phases and in the way they predict
physical properties of the phases of Pu.
For example, the approaches using orbital polarization, spin polarization, or LDA+$U$
\cite{soderlind-2,bouchet,price,savrasov-LDAU,soderlind-4}
predict a magnetically ordered ground state of $\delta$-Pu, albeit with
spin and orbital moments that almost cancel.
This is in contradiction to the observed temperature independent magnetic susceptibility of the
$\delta$-phase\cite{pudata}. There is in fact no evidence of magnetic moments
in the $\delta$-phase, either ordered or disordered.
Further, the works of
Refs.~\onlinecite{soderlind}, \onlinecite{eriksson} and \onlinecite{soderlind-3}
indicate that in the $\alpha$-phase, the 5f electrons are delocalized and
band-like. According to this picture,
the 5f electrons are responsible for the low-symmetry monoclinic structure of $\alpha$-Pu, which
is correctly predicted from GGA calculations\cite{soderlind-3},
whereas in the $\delta$-phase they are more localized.
This band picture of $\alpha$-Pu is very different from the one put forth in
Ref.~\onlinecite{savrasov-DMFT}
where, from calculations using an LDA+$U$+DMFT approach,
it is argued that correlation effects are very important
not only in the $\delta$-phase but also in the $\alpha$-phase.
Unfortunately, structural effects were not addressed in the latter
study, and with this model one has not been able to assess how
the effect of correlations influence the theoretical
description of the structural properties of Pu.

In order to resolve the controversies regarding the theoretical
models of the electronic structure of the $\delta$-phase it
becomes very important to test their ability to reproduce all
types of  experimental data. In this report we present a theory
that, unlike previous theories, is found to be consistent with the
cohesive, structural, magnetic, and electronic properties of
$\delta$-Pu.

The basis of our model is a partitioning of the electron density
into localized and delocalized parts, minimizing the total energy
(including a correlation energy associated with localization) with
respect to the partitioning. The total energy for Pu becomes
minimized for an atomic 5f$^4$ configuration\cite{eriksson} with
roughly one 5f-electron in a delocalized Bloch state. The four
localized 5f electrons couple into a singlet, a true many-body
effect that can never be accounted for within a single-particle
picture. As will be discussed below a model based entirely on
localized electrons is too simple, and it can not reproduce the
observed photo emission spectrum of $\delta$-Pu. We will show that
the hybridization between the conduction band states and the
localized f-states is the key interaction needed to procude an
accurate theoretical spectrum. This hybridization induces
broadening of the 5f$^4$ multiplet, with a spectral function, over
the valence binding energies $\epsilon$, with the following shape:

\begin{eqnarray}
\rho(\epsilon)=\frac{N}{\pi} \frac{V}{(\epsilon-e_{\rm f})^2+V^2}
\label{eq:one},
\end{eqnarray}
where $N$ is the 5f occupation number,
$V$ the hybridization matrix element and
$e_{\rm f}$ is the energy of the localized 5f singlet.
The hybridization matrix element is calculated from the width of the 5f-resonance
which is proportional to the 5f electron wavefunction at the muffin-tin sphere.
In a symmetric Hubbard model,
$E_{\rm F} - e_{\rm f}$ would equal $U/2$, where $U$ is the intra-atomic f-electron Coulomb interaction, usually
assumed to be 4 eV for Pu.
Here, $E_{\rm F} - e_{\rm f}$ is chosen to be 1.2 eV, a smaller value than the symmetric Hubbard model would
suggest. However, our value is entirely consistent with the fact that spin- and orbital couplings within the 5f shell
are important in this system, since
in the experimental photoemission spectrum, these  couplings
will influence the positions of the lower and upper Hubbard bands.

In addition to a broadening of the 5f-level, the hybridization matrix elements cause a reduction in
the spectral weight corresponding to
\begin{eqnarray}
\Delta N=- \frac{V^2}{(\epsilon-\epsilon_{\rm f})^2+V^2}
\label{eq:two}.
\end{eqnarray}
Hence the hybridization results in a broadening and a shift of the spectral weight.

Combined with the calculated relativistic cross sections of the photoemission
process\cite{cross}
a Fermi-liquid like Lorentzian broadening for the
photohole that scales quadratically with the binding energy,
a Gaussian broadening of 60
meV for the instrument resolution, a Shirely background to remove
the scattered electron background and a 15~K Fermi function, we
obtain a theoretical photoemission spectrum that is compared to our
experimental data\cite{PES-footnote} for the $\delta$-phase\cite{Inset-footnote} in Fig.1.
Several experimental valence photoemission spectra for $\delta$-Pu
have been published
previously\cite{naegele,arko,gouder,havela,terry}. We note that
the general features of our experimental spectrum are very similar
to the thin film Pu data reported in Ref.~\onlinecite{havela} but
it is different from the spectrum of Ref.~\onlinecite{terry},
probably due to oxidation of that sample and the temperature used
in that study.

Our calculated spectrum shows a very close similarity to the
experimental one. It reproduces the peak close to $E_{\rm F}$, the
broader feature between 0.5 and 2 eV binding energy and the valley
between the two features. The peak at $E_{\rm F}$ is the result of
hybridized 5f--6d electron states that form naturally in our
theory. We note that the dip in the theoretical spectrum at 0.5 eV
binding energy is more pronounced than what is observed in the
experiment, but this may be a result of the details of the
scattered electron background removal function. The calculated
peak at $E_{\rm F}$  is also slightly shifted (50 meV) to higher
binding energy compared to the experiment. This discrepancy is
within the calculational error of our method.

\begin{figure}
\vspace*{4mm}
\includegraphics[scale=0.5]{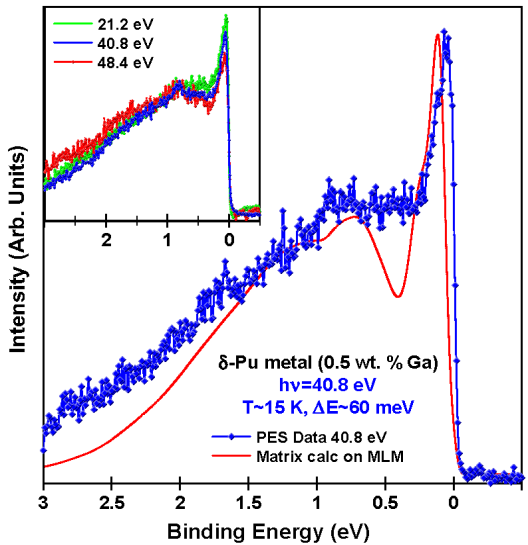}
\caption{\label{fig}Experimental and theoretical photoemission
spectrum for $\delta$-Pu. The experimental spectrum is given as
dots connected with lines (blue), whereas the theoretical spectra
is given as a solid line. The inset shows how the experimental
spectrum varies with photon energy.}
\end{figure}

The fact that the peak at $E_{\rm F}$ consists of hybridized 5f and valence band
states
is consistent with our experimental data when we vary the photon
energy (see the inset in the figure). By increasing the photon
energy the cross section of the 5f states is increased and the
cross section of the 6d states is lowered. Hence any feature in
the experimental spectrum that contains 6d character would reduce
its intensity over the 5f component and in the experimental data
this does indeed occur for the peak at the Fermi
level.\cite{Inset-footnote}
An analysis based purely on localized electron states would fail in describing this behavior.

In order to assess the quality of our theory,
it is important to compare its ability to reproduce the experimental spectrum to other theoretical models.
A non-spin polarised energy band calculation
reproduces the experimental peak at $E_F$, but fails in describing
the observed features at binding energies higher than 0.5 eV.
A spin polarized calculation, either in a ferromagnetic, anti-ferromagnetic configuration\cite{soderlind-4},
or disordered local moment configuration\cite{niklasson}
improves the agreement with the experimnetal photo electron spectrum,
but the areement is not as good as that given by the theory proposed here.
The calcualtion based on the LDA+U approximation\cite{savrasov-LDAU}
results in a dominating broad peak centered at 1.5 eV binding energy, which is in disagreement with experiment.
Finally, the calculated spectrum based on DMFT has a $\sim$ 1 eV broad peak at $E_F$ with very small spectral
features at higher binding energies\cite{savrasov-DMFT}, resulting in a rather poor agreement with experiment.

To summarise, we have presented a novel theoretical model of the
electronic structure and the photoemission spectrum of
$\delta$-Pu. The observed features of the experimental spectrum
are well reproduced by our theory, whereas
all previous calculations \cite{savrasov-DMFT,savrasov-LDAU,petit} do not
reproduce this spectrum with the same accuracy.
A key element of our model is
to recognize the importance of the many-body spin- and orbital
couplings within the 5f$^4$ singlet. Of these, the spin pairing energy
is the most important (Hund's first rule).
Allowing for spin polarization in an electronic structure calculation, as done in
Refs.~\onlinecite{soderlind-4} and ~\onlinecite{niklasson}, does incorporate some of these interactions and the calculated
photo emission spectrum based on this theory reproduces some of the experimental features
of $\delta$-Pu, although not with as good accuracy as that given by the present theory.

{\bf Acknowledgment} Work performed under the auspices of The U.S.
Department of Energy. O.E. Acknowledges support from the Swedish
Research Council (VR), the G\"oran Gustafsson foundation and the
foundation for strategic research (SSF). A.D. acknowledges support
from the IHP Marie Curie Fellowships program.


\begin{thebibliography}{10}
\bibitem{PES-footnote}
The experimental spectrum was recorded on Ga-stabilized
$\delta$-Pu at a temperature of 15 K with an experimental
resolution of 60 meV. The measurements were made at a pressure of
5x10$^-11$ Torr. The sample was cleaned by means of laser ablation
while held at low temperature resulting in an oxygen-free
$\delta$-Pu metal spectrum.
\bibitem{Inset-footnote}
The data are from a $\delta$-Pu alloy of Pu with 0.5 weight
percent Ga. This alloy was formed via a wrought process and had a
grain size of approximately 45 microns. The total amount of
metallic impurities in the material is less than 1000 ppm. The
data have been background subtracted and in the inset, normalized
in intensity to the 5f peak at 1 eV below the Fermi level.
\end{thebibliography}
\end{document}